# Demystifying algorithmic complexities and geometric review of the 'h'-index

by Kaushik Ghosh & Mayukh Mukhopadhyay

**Authors details:**

Kaushik Ghosh (Corresponding Author)
BE (IT), West Bengal University of Technology
MBA, Vinod Gupta School of Management (VGSoM)
Indian Institute of Technology (IIT) Kharagpur
Junior Works Manager
Ordnance Factory Board, Ministry of Defence, Govt of India
Professional Member, Association for Computing Machinery (Membership number 8912100)
West Bengal, India
Present and Permanent Address: 115/61, Palash Sarani, Bhadrakali, Hooghly, West Bengal, Pin: 712232
e-mail: kaushikghosh@iitkgp.ac.in

Mayukh Mukhopadhyay
MBA, Vinod Gupta School of Management (VGSoM)
Indian Institute of Technology (IIT) Kharagpur
Assistant Consultant
Tata Consultancy Services
Kolkata
West Bengal, India
Present and Permanent Address: E-3 Shree Kunja
2/1, Panchanantola Road
Sukchar
Kolkata - 700115
e-mail: mayukhmukhopadhyay@iitkgp.ac.in

**Declaration:**

- The authors declare that there is no conflict of interest with anyone.
- No fund has been received from anywhere to conduct this research.

## Abstract:

The current discourse delves into the effectiveness of h-index[1] as an author level metric. It further reviews and explains the algorithmic complexity of calculating h-index through algebraic method. To conduct the algebraic analysis propositional algebra, algorithm and coding techniques have been used. Some use cases have been identified with a finite data set/set of array to demonstrate the coding techniques and for further analysis. Finally, the explanation and calculative complexities to determine the index have been further simplified through geometric method of calculating the h-index using the similar use cases that was used for coding. It is concluded that determination of the h-index using Euclidean geometric method with Cartesian frame of reference provides a through and visual clarification. Finally, a set of postulates has been proposed at the end of the paper, based on the case studies.

**MSC Classification Codes**: 51N20, 52B55, 65Y04, 68Q25, 68W40

---

[1] **Defining H – index:** 'H'-index/number stands for Hirsch index or Hirsch number. The h-index is defined as the author-level metric that attempts to measure both the productivity and the citation impact of the publication of the scientist or the scholar. The index was suggested in 2005 by Jorge E. Hirsch, a physicist at UC San Diego, as a tool for determining theoretical physicists' relative quality and is sometimes called the **Hirsch index or Hirsch number**.

## 1.   Introduction - H-index - Definition and explanation:

JE Hirsch (2005) in his Proceedings of the National Academy of Sciences (PNAS) 102:16569-16572[2] (An index to quantify an individual's scientific research output), stated the following:

*The publication record of an individual and the citation record clearly are data that contain useful information. That information includes the number ($N_p$) of papers published over n years, the number of citations ($N^c_j$) for each paper (j), the journals where the papers were published, their impact parameter, etc.*

*This large amount of information will be evaluated with different criteria by different people. Here, I would like to propose a single number, the ''h index,'' as a particularly simple and useful way to characterize the scientific output of a researcher.*

**A scientist has index h *if* h *of his or her* $N_p$ *papers have at least* h *citations each and the other (*$N_p$ *− h) papers have ≤h citations each.***

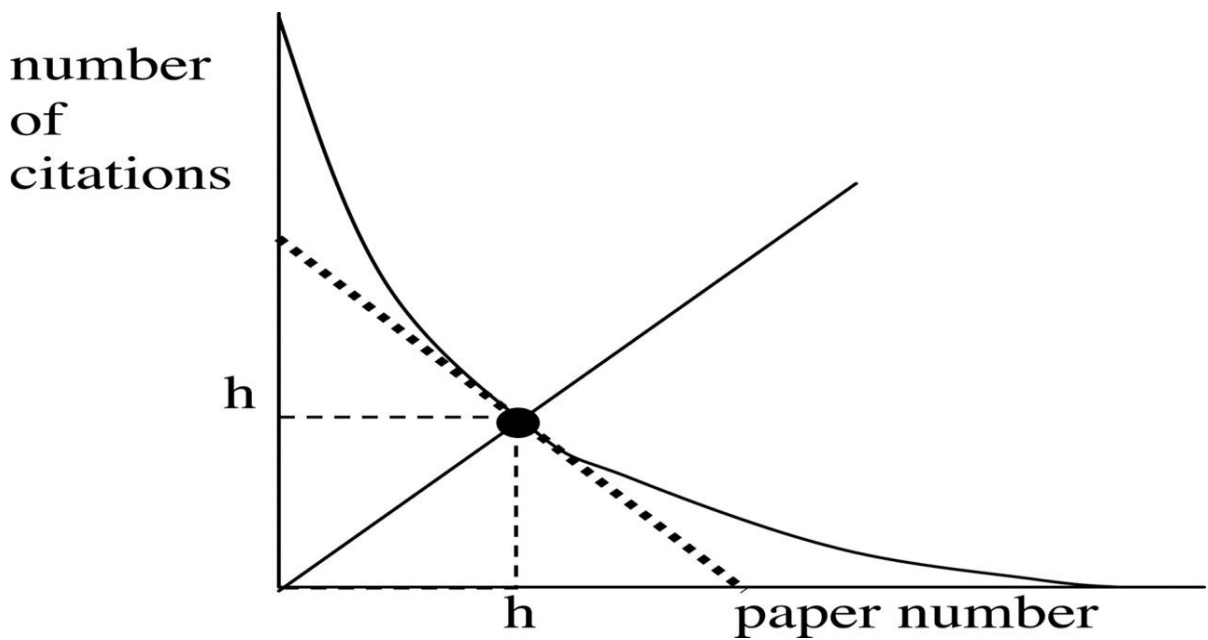

Figure 1: Source: Footnote 2.

---

S. Alonso, F.J. Cabrerizo, E. Herrera-Viedma, F. Herrera, h-index: A Review Focused in its Variants, Computation and Standardization for Different Scientific Fields. Journal of Informetrics 3:4 (2009) 273-289, depicted h-index by the following figure:

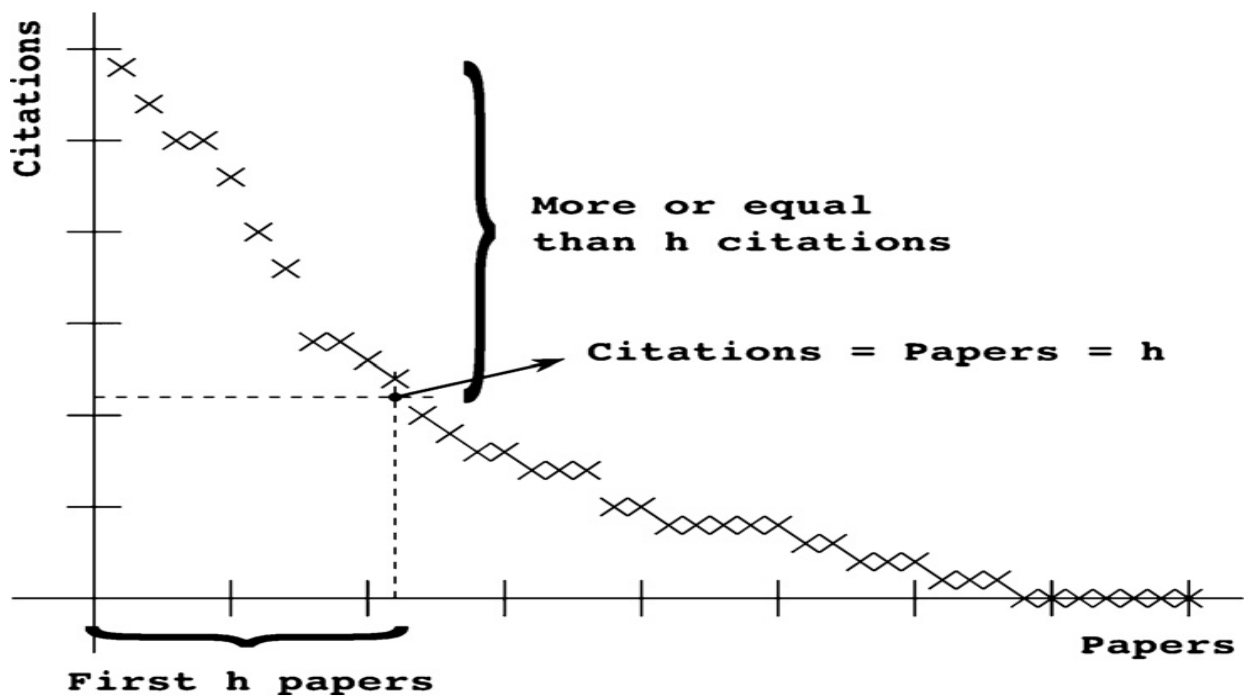

Figure 2: Graphical interpretation of the h-index ; Source[3]

**1.1** H-index considers the following two parameters –

1. **Quantity of productivity** – which is given by the Numbers of papers
2. **Quality of productivity** – which is given by the Number of citations

In other words, the index is based on the set of the scientist's most cited papers and the number of citations that they have received in other publications. H-index takes into account two dimensions of productivity, quality and quantity.

Clarivate[4] in their article "Web of Science: h-index information" identifies the pivotal role played by the h-index as: *The h-index is based on a list of*

---

[3] S. Alonso et al. / Journal of Informetrics 3 (2009) 273–289, doi:10.1016/j.joi.2009.04.001
[4] Clarivate is a company formed in 2016, following the acquisition of Thomson Reuters' Intellectual Property and Science Business by Onex Corporation and Baring Private Equity Asia. Clarivate owns and operates a collection of subscription-based services focused largely on analytics, including scientific and academic research, patent intelligence and compliance

*publications ranked in descending order by the Times Cited. The value of h is equal to the number of papers (N) in the list that have N or more citations. This metric is useful because it discounts the disproportionate weight of highly cited papers or papers that have not yet been cited.*

The measure of H-index is the largest number of publications such that the number of publications have at least the same number of citations. As a useful index to characterize the scientific output of a researcher. As previously defined, the index determines both quality and quantity of a researcher.

Lutz Bornmann by Hans-Dieter Daniel in their text "What do we know about the h index?[5]" stated that The *h*-index correlates with obvious success indicators such as winning the Nobel Prize, being accepted for research fellowships and holding positions at top universities.

## 2. Advantages of h-index[6]:

 i. H-index combines a measure of quantity (number of publications) and impact/quality of each publication (given by citations).
 ii. This characterizes the scientific output of a researcher with objectivity of research and thus can be considered as a level playing field among researchers.
 iii. It performs better than other single-number criteria commonly used to evaluate the scientific output of a researcher (impact factor, total number of documents, total number of citations, citation per paper rate and number of highly cited papers).
 iv. The h-index can be easily deetermined by anyone with access to the Web of Science data.
 v. H-index is easy to understand.

## 3. H-index from an algebraic point of view:

**To conduct the algebraic analysis propositional algebra, algorithm and coding techniques have been used. Some pre-defined use cases have been identified and used as a finite data set/set of array to demonstrate the coding techniques and for further analysis.**

**Initially the research views the determination of h-index as an Optimization problem with a maximization-minimisation constraint in**

---

standards, pharmaceutical and biotech intelligence trademark, domain and brand protection. The services include Web of Science.

[5] Journal of the American Society for Information Science and Technology, Volume 58, Issue 9, July 2007, Link: https://doi.org/10.1002/asi.20609

[6] Costas R, Bordons M (2007) Advantages, limitations and its relation with other bibliometric indacators at the micro level. Journal of Informetrics 1(3):193-203, doi: 10.1016/j.joi.2007.02.001

**the problem. To further elucidate,** this problem has the following mathematical form:

The largest (or smallest) value of *f(x)* when $a \leq x \leq b$. This constraint shows that both *x* & *f(x)* shall have finite set of values.

*x* & *f(x)* are respectively the number serial of papers/journals and the number of citations received. Unlike other mathematical functions, no specific mathematical or numerical relationship can be identified or established for all sets of *x* & *f(x)*. *x* being a serial integer & *f(x)* being statistical data, which too is an integer. More often than not, value of *f(x)* appears as random integer numbers with varying digits. When *x* & *f(x)* are plotted in a graph in an cartesian coordinate graph, some relationship may be established between *x* & *f(x)*, especially when they are arranged in increasing and decreasing order. This feature is explained by the following proposition:

### 3.1 The maxima-minima constraint finally leads to a win-win situation with zero trade-off:

As discussed H-index is an optimisation problem, **with a maximization-minimisation constraint**. H-index a game to determine a win-win situation of papers and citations. Here we maximise the number of papers/journals for which at least a minimum number of citations are available for each paper. H-index does not compromise quality of the papers with the quantity or vice-versa. It is actually the culmination of optimality problem of quality and quantity; in both ways the optimal productivity is maximised in the h-index.

Maxima and minima refers to, **at most** how many journals/papers are there which have been cited **at least** n times and this n should not be less than the journal number.

The logical mathematical/algebraic model that represents the h-index is as follows,

Let us consider, cumulative (and increasing) number of papers/journals be x and citations arranged in the decreasing order, be y.

For all natural positive integers, i.e. ($\forall$ x, y $\in$ N or $\forall$ x, y $\in$ Z+)

x = |F(y)|; $\Leftrightarrow$ (iff) $\forall$ (x = |F(y)|) $\in$ N and $\forall$ (x = |F(y)|) $\in$ Z+

$\lfloor$ x $\rfloor$ = |F(y)|; $\Leftrightarrow$ $\forall$ (x = |F(y)|) $\in$ N, $\forall$ (x = |F(y)|) $\notin$ Z and $\forall$ (x = |F(y)|) $\notin$ Z+

In other words,

h-index or *h*-index (*f*) = max { i $\in$ N : f(i) > i} $\forall$ i $\in$ Z+

where,

i is a function increasing by 1 for each corresponding y (number of citations), ideally sorted or unsorted, starting from 1 to n

So, i = 1,2,3,.......n

In mathematical notations,

h-index function is defined by, $h\text{-index}(f) = \max\{i \in N : f(i) > i\} \ \forall \ i \in Z^+$

## 3.2  Upper limit of h-index:

In simple words, the H-index is always less than or equals to total numbers of papers published. So h-index $(f)_{max} = n$, n is the number of papers published.

The index can also be applied to the productivity and impact of a scholarly journal as well as a group of scientists, such as a department or university or country.

Critique: Though Wikipedia defined the H-index as an "**h-index** is an author-level metric that measures both the productivity and citation impact of the publications of a scientist or scholar"; the same is also frequently ben used to determine Journal Impact Factor as a deterministic measure for Journal quality by Clarivate[7].

**3.3  Calculating the H-index** – H-index is always less than or equals to total numbers of papers published

For example, consider a Researcher has published total 10 papers. The h-index cannot be more than 10.

## 4.  Simple algorithmic approach for finding the H – index:

### 4.1  An algorithm to determine the h-index is proposed below:

1. Sort the citation array either in ascending order or descending order.
2. Increase the paper number lowest serial number paper to the highest serial number paper.

---

[7] **Clarivate** is a company formed in 2016, following the acquisition of Thomson Reuters' Intellectual Property and Science Business by Onex Corporation and Baring Private Equity Asia. On May 13, 2019, Clarivate merged with Churchill Capital Corp and was publicly listed on the New York Stock Exchange.
Clarivate owns and operates a collection of subscription-based services focused largely on analytics, including scientific and academic research, patent intelligence and compliance standards, pharmaceutical and biotech intelligence trademark, domain and brand protection.

3. Run the iteration as Sl. No. 3 till the paper number falls short/less than the corresponding value of citation in the citation array.
4. Whenever the paper number equals with or just overreaches the corresponding number of citation stop the iteration. Note the corresponding paper number.
5. Declare the paper number (in case of equal with the corresponding citation number) or floor value, i.e. next lower integer of the number (in case of overreaching) as the h-index.
6. The derived serial number (of the paper) as is the Sl. No. 4 is the H-index. The index should always be less than or equal to total number off cited paper.

## 4.2 H-index: The coding perspective:

The following Python code has been used for calculating H-Index

```python
def H_index(citations):

    # sorting in ascending order
    citations.sort()

    # iterating over the list
    for i, cited in enumerate(citations):

        # finding current result
        result = len(citations) - i

        # if result is less than or equal
        # to cited then return result
        if result <= cited:
            return result

    return 0

# Entering the array of number of citations in descending order
citation = [10, 9, 8, 8, 7, 5, 4, 3, 2, 1, 1]

# calling the function

print(H_index(citation))
```

**Output is given as**

## 4.3 Analysis of the Algorithm:

Sorting

The Python list sort() has been using the **Timsort** algorithm since version 2.3.

This algorithm has a runtime complexity of *O(nlogn)*.

So, the citations.sort() has a time complexity of *O(nlogn)*

Now, since, result = len(citations) – i has linear time complexity and the complexity can be maximum n, i.e., the number of the elements in the input array of number of citations, so

**Here for the sorting algorithm the time complexity is calculated as:**

**Time Complexity:** O(nlogn + n)
**Space Complexity:** O(1)

**Limitations of H – Index:**

1. Different fields of researchers can have different citation patterns and behaviour. For an example, a less experienced author of Humanities may have less number of citations as compared to a less experienced author of Technology.
2. Researchers with varying research experiences and having different fields of research are therefore not comparable with their varying citation patterns. Experienced researcher will have high h-index as compare to less experienced researcher.
3. Therefore, h-index alone cannot determine quality of research across the fields of research.
4. H-index value depends on the database in which it is stored and it is not platform independent, i.e., it may vary for the different platform.

**4.4  R-code for determining h-index:**

```
h_index = function(cites) {
  if(max(cites) == 0) return(0) # assuming this is reasonable
  cites = cites[order(cites, decreasing = TRUE)]
  tail(which(cites >= seq_along(cites)), 1)
}
a1 = c(10, 9, 8, 8, 7, 5, 4, 3, 2, 1, 1)
a2 = c(10, 9, 7, 3, 2, 1, 1)
a3 = c(4, 3, 2, 1)
a4 = c(400, 300, 200, 2)
a5 = c(700, 600, 8, 7, 7, 6)
h_index(an)
```

# here an represents n = 1 to 6 all integer numbers, i.e., a1, a2,...a6

Output is given as below:

# a1 is 5; # a2 is 3; # a3 is 2; # a4 is 3; # a5 is 6

On the contrary, if we use the sorted array in an ascending order, the output remains the same for the above code snippet.

A6 = c (1, 1, 2, 3, 4, 5, 7, 8, 8, 9, 10)

5. **Geometric analysis and determination of the h-index:**

In Cartesian co-ordinate geometry this H-index is represented as the point of intersection between the Citations Coordinate and Journal number (which constitutes the cumulative total count incremented by 1) coordinate. In the present discourse, the Citations Coordinate represents the Y-Axis and Journal number coordinate is represented in the X-Axis.

This situation is true if and only if the point of intersection of the two lines is an integer. In case the point of intersection of the two lines is not an integer then the lower limit of thee integer, i.e., the next smaller integer number will always represent the h-index. In this way the h-index suffers from inaccuracy. A better representation of h-index requires to be in decimal points. The rationale used for making the h-index an integer is the indivisibility of number of journals, i.e., the number of journals cannot be expressed in fraction/floating point.

Let us consider, cumulative (and increasing) number of papers/journals be x and citations arranged in the decreasing order, be y.

For all natural positive integers, ($\forall$ x, y $\in$ N and $\forall$ x, y $\in$ Z+)

The Cartesian geometrical model that represents the h-index is,

Point of intersection between positively sloped, cumulative number of papers/journals and negatively sloped, citations arranged in the decreasing/descending order.

If the Point of intersection is a positive integer, then h-index = length of the perpendicular drawn from the point of intersection on the either axis, i.e., x or y axis.

If the Point of intersection is a positive non-integer, i.e., a fraction then the h-index is calculated as next lower integer, i.e. the floor function to the length of the perpendicular drawn from the point of intersection on the either axis, i.e., x or y axis.

## 5.1 Case wise geometric analysis and calculation of the h-index:

Geometric significance of the h-index, where Journal number line (always a straight line) and Citation(s) represented by a straight line or a nearly straight line or a curve. This two lines may or may not intersect each other.

In the next part the Case-wise determination, analysis and geometrical significance of h-index is described.

### 5.1.1 Case a1:

Case a1 represents h-index, where Journal number and Citation(s) respectively represented by a straight lines and a nearly straight line that intersect each other at only one point.

The case a1 have been analysed as below:

| Case a1 | | | | | | | | | | | |
|---|---|---|---|---|---|---|---|---|---|---|---|
| Sl. No. | 1 | 2 | 3 | 4 | 5 | 6 | 7 | 8 | 9 | 10 | 11 |
| Journal number (x-axis) | 1 | 2 | 3 | 4 | 5 | 6 | 7 | 8 | 9 | 10 | 11 |
| Citation(s) (in descending order) (y-axis) | 10 | 9 | 8 | 8 | 7 | 5 | 4 | 3 | 2 | 1 | 1 |

The above points are represented in a Cartesian coordinate system

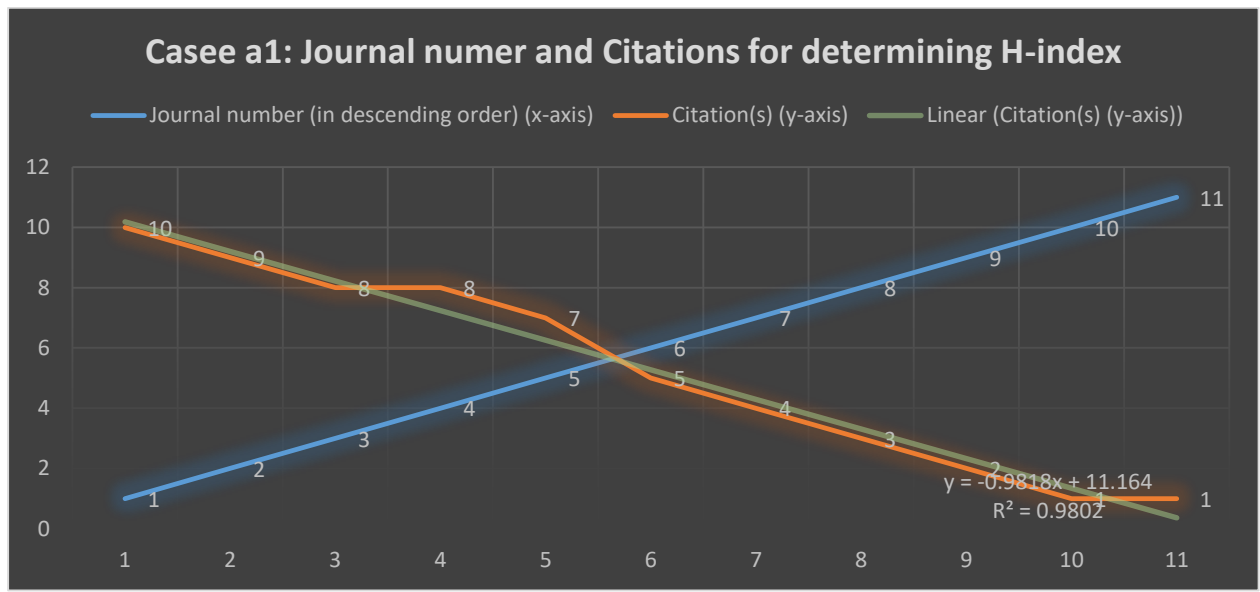

| Citations Coordinate (brown line) | x | 1 | 2 | 3 | 4 | 5 | 6 | 7 | 8 | 9 | 10 | 11 | Slope | Intercept |
|---|---|---|---|---|---|---|---|---|---|---|---|---|---|---|
|  | y | 10 | 9 | 8 | 8 | 7 | 5 | 4 | 3 | 2 | 1 | 1 | -0.98 | 11.16 |
| Journal Number Coordinate (blue line) | x | 1 | 2 | 3 | 4 | 5 | 6 | 7 | 8 | 9 | 10 | 11 | Slope | Intercept |
|  | y | 1 | 2 | 3 | 4 | 5 | 6 | 7 | 8 | 9 | 10 | 11 | 1 | 0 |

Now it is evident from the figure that

Using the y = mx + c, where m= slope, c = intercept

Here, the Citation(s) line in the graph, which is negatively sloped is coloured in brown.

The Journal Number Line in the graph, which is positively sloped and a straight-line, is coloured in blue.

The Citations Trend-line which is negatively sloped and a straight-line, is olive-coloured.

Citation(s) Line Equation is y = -0.98182x + 11.164

Journal Number Line Equation is y = x

Trendline Equation is y = -0.9818x + 11.164

Y-Axis intercept value of the Trendline (olive colour line) is 11.16364

It clearly shows that the Trendline equation is nearly the same as the Citation equation.

Putting y = x in both the equations we get,

The two lines intersect at (5.633206, 5.633206)

Table: Determination of the h-index through Cartesian geometry:

|  | X = |
|---|---|
| Citation & Journal point of intersection co-ordinate | 5.633206 |
| Trendline & Journal point of intersection co-ordinate | 5.633263 |
| Taking the lower limit (floor) of the calculated x, i.e. ⌊ x ⌋ = | 5 |
| Geometrically Calculated h-index (same as the original h-index) | 5 |
| Y-Axis intercept value of the citations line (brown line) | 11.16364 |

### 5.1.2 Case a2:

The case a2 has been analysed as below:

| Sl. No. | 1 | 2 | 3 | 4 | 5 | 6 | 7 |
|---|---|---|---|---|---|---|---|
| Journal number (x-axis) | 1 | 2 | 3 | 4 | 5 | 6 | 7 |
| Citation(s) (in descending order) (y-axis) | 10 | 9 | 7 | 3 | 2 | 1 | 1 |

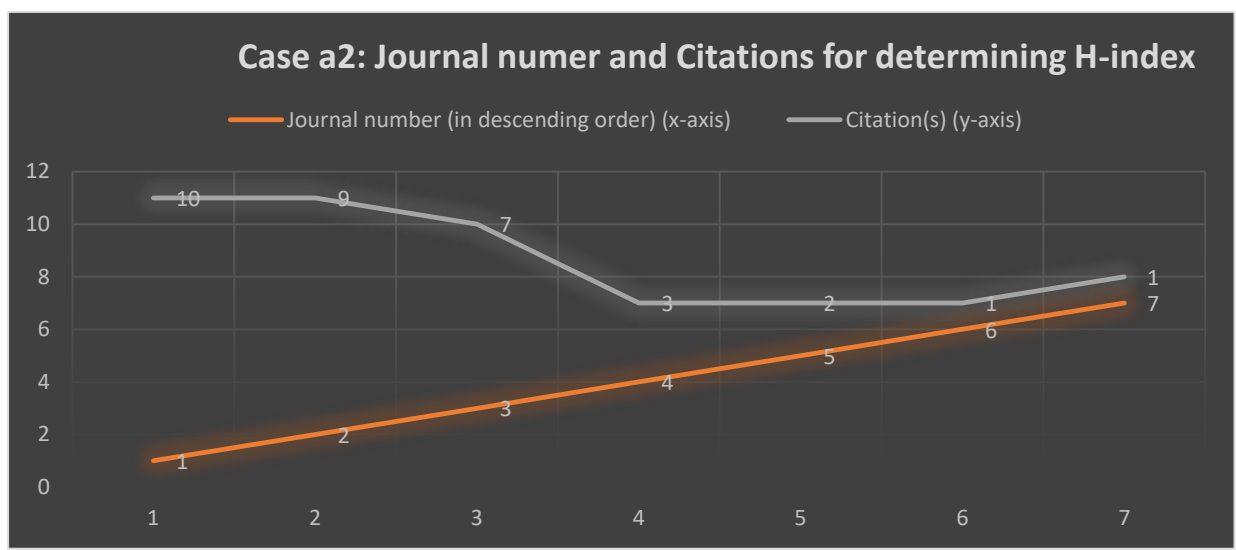

| | | x | 1 | 2 | 3 | 4 | 5 | 6 | 7 | Since the citation line is not ideally or nearly a straight line and it has curvilinear nature, the slope is not drawn. | |
|---|---|---|---|---|---|---|---|---|---|---|---|
| Citations Coordinates (grey line) (q) | | y | 10 | 9 | 7 | 3 | 2 | 1 | 1 | | |
| Journal Number Coordinates (brown line) (q) | | x | 1 | 2 | 3 | 4 | 5 | 6 | 7 | Slope | Intercept |
| | | y | 1 | 2 | 3 | 4 | 5 | 6 | 7 | 1 | 0 |
| Perpendicular/Euclidean distance between (p) and (q) | | | 9 | 7 | 4 | 1 | 3 | 5 | 6 | Minimum of all perpendicular distances = Min(9,7,4,1,3,5,6) = 1 | |
| Calculated Geometric h-index = 3 | | | | | | | | | | | |

For Two dimensions: In the Euclidean plane, let point p have Cartesian coordinates (p1,p2) and let point q have coordinates (q1, q2). Then the distance between points p and q is given by: d (p, q) = $\sqrt{\{(q1-p1)^2 + (q1-p1)^2\}}$.

Case 2 is an ideal example where the citation line and journal number line (or curve) does not intersect each other at any point. Since the citation line is not ideally or nearly a straight line and it has curvilinear nature, the slope for the journal number line (or curve, i.e., the grey line/curve) is not drawn.

Had the minimum distance been zero the two lines could have intersected each other, The h-index is therefore calculated as one less than the journal number serial, corresponding to the minimum perpendicular distance, i.e., 1. This is the point at which the journal number crosses the number of citations, this satisfies the maxima-minima criteria as stated earlier.

Here, Minimum of all perpendicular distances = Min (9,7,4,1,3,5,6) = 1 (marked in yellow), corresponding journal number is 4 1 (marked in grey) so one less than that journal number is 3 (marked in green).

Therefore, calculated h-index = 3.

**5.1.3 Case a3:**

The case a3 has been analysed as below:

| Case a3 | | | | |
|---|---|---|---|---|
| Sl. No. | 1 | 2 | 3 | 4 |
| Journal number (in descending order) (x-axis) | 1 | 2 | 3 | 4 |
| Citation(s) (y-axis) | 4 | 3 | 2 | 1 |

| Citations Coordinates (grey line) (q) | x | 1 | 2 | 3 | 4 | Slope | Intercept |
|---|---|---|---|---|---|---|---|
| | y | 4 | 3 | 2 | 1 | -1 | 5 |
| Journal Number Coordinates (brown line) (q) | x | 1 | 2 | 3 | 4 | Slope | Intercept |
| | y | 1 | 2 | 3 | 4 | 1 | 0 |

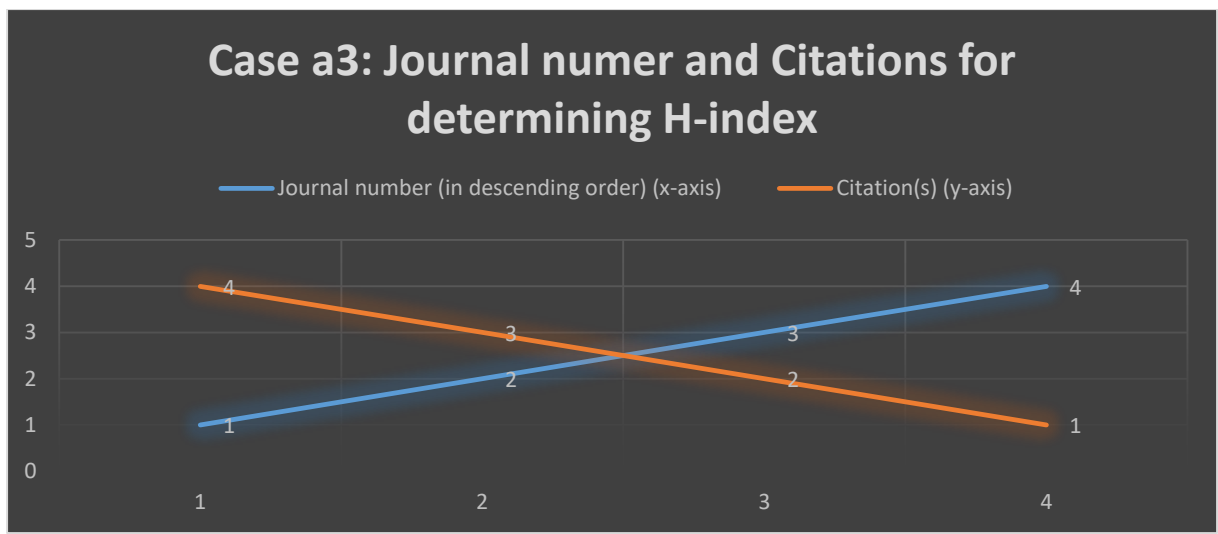

According to the above figure and above tables,

Citation(s) Line Equation is y = -x + 5

Journal Number Line Equation is y = x

Therefore, 2x = 5 or, x = 2.5, y = 2.5

The two lines intersect at (2.5, 2.5)

Floor of x, i.e., $\lfloor x \rfloor = 2$

Therefore, calculated h-index = 3.

### 5.1.4 Case a4:

The case a4 has been analysed as below:

| Case a4 | | | | |
|---|---|---|---|---|
| Sl. No. | 1 | 2 | 3 | 4 |
| Journal number (in descending order) (x-axis) | 1 | 2 | 3 | 4 |
| Citation(s) (y-axis) | 400 | 300 | 200 | 2 |

| | | x | 1 | 2 | 3 | 4 | Since the citation line is not ideally or nearly a straight line and it has curvilinear nature, the slope is not drawn. | |
|---|---|---|---|---|---|---|---|---|
| Citations Coordinates (brown line) (q) | | y | 400 | 300 | 200 | 2 | | |
| Journal Number Coordinates (blue line) (p) | | x | 1 | 2 | 3 | 4 | Slope | Intercept |
| | | y | 1 | 2 | 3 | 4 | 1 | 0 |
| Perpendicular/Euclidean distance between (p) and (q) | | | 399 | 298 | 197 | 2 | Minimum of all perpendicular distances = Min(399,298,197,2) = 2 | |

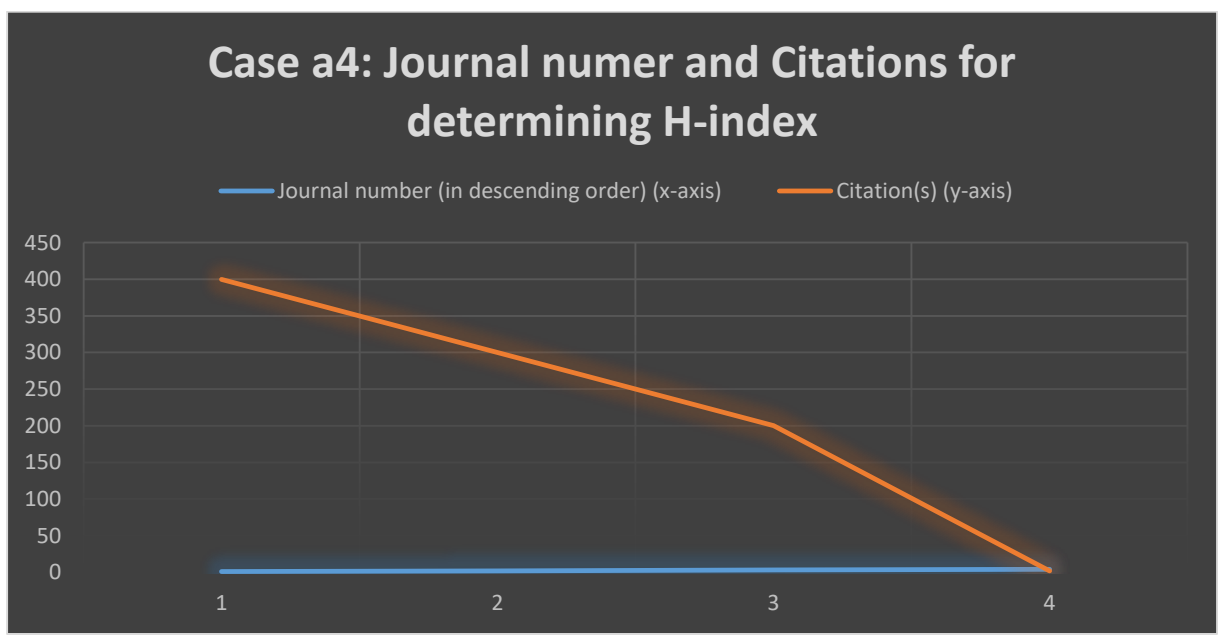

Case a4: Journal numer and Citations for determining H-index

Apparently, from the above table and the picture, the h-index is calculated as 2. Since the table suggests that Minimum of all perpendicular distances between the two lines = Min (399,298,197,2) = 2 and the picture suggests that the two lines intersect at x = (4,2). Now since 4>2 so, at the point of intersection, Journal Number Coordinate (blue line) is greater than Citations Coordinates (brown line). This itself violates the h-index criterion, i.e.,

Journal serial number (p) must be greater or equal to the no. of cited papers (q), i.e., p>q.

In the previous case, i.e., at the point x = 3 the journal number equals to 200. & 3 ≮ 200.

So, determined h-index = 3.

### 5.1.5 Case a5:

Now we move on to the case no. 5 and final case of our use cases,

| Case a5 | | | | | | |
|---|---|---|---|---|---|---|
| Sl. No. | 1 | 2 | 3 | 4 | 5 | 6 |
| Journal number (in descending order) (x-axis) | 1 | 2 | 3 | 4 | 5 | 6 |
| Citation(s) (y-axis) | 700 | 600 | 8 | 7 | 7 | 6 |

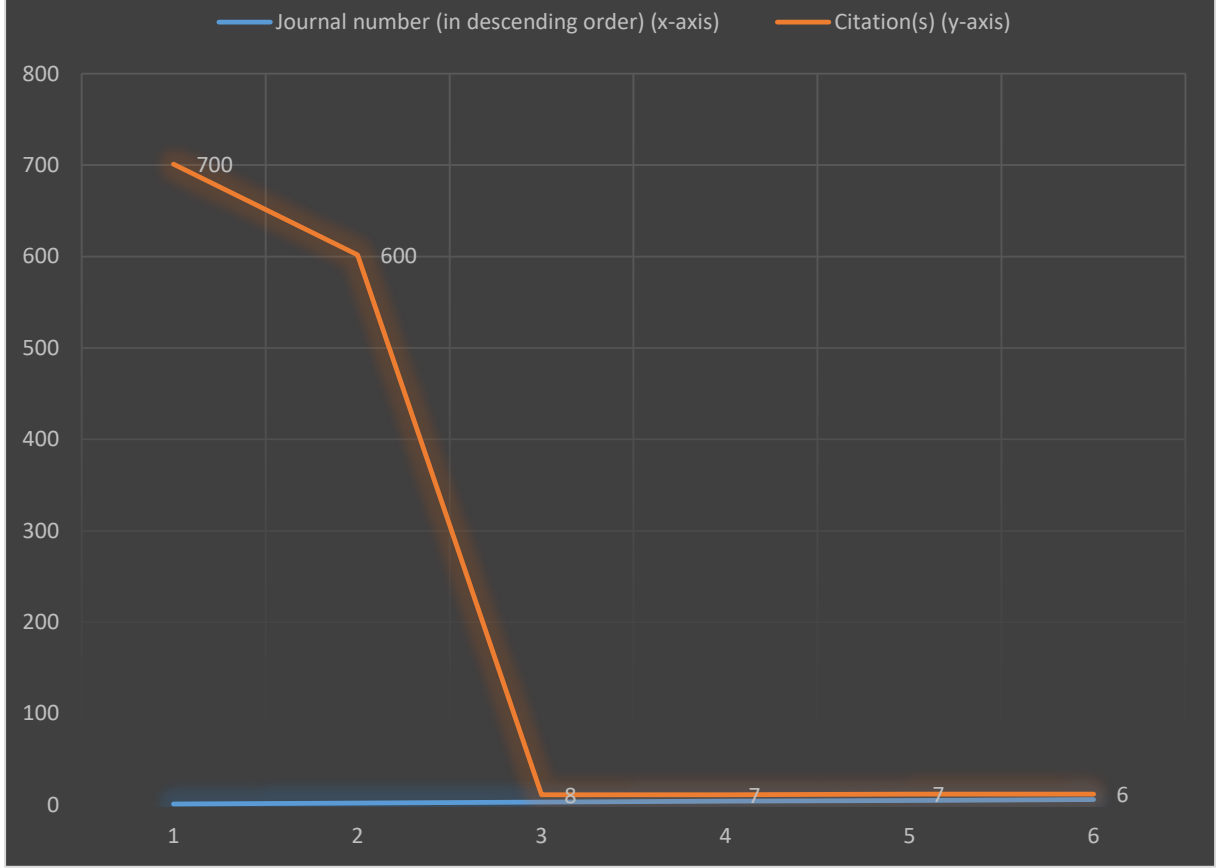

Without analysing the case numerically, we see that the two lines of the points (6,6) intersects at each other.

So, the determined h-index here is 6.

# 6. From the above experience of geometric mapping of the use cases, following postulates are formulated:

The postulates for geometrically determining the h-index:

Considering the Journal Number serial line is increasing and positive sloping and drawn along x-coordinate. Number of Citations line is arranged in decreasing order with increasing journal number and Number of Citations line is negative sloping and drawn along y-coordinate.

i. When the two lines, i.e., Journal Number serial line and Number of Citations lines intersect each other and the value of the coordinates of the points of intersection are integers, then at the point of intersection:
   a. If value of the x-coordinate equals to the value of the y-coordinate, then the h-index is equal to that integer, i.e., h-index = x = y.
   b. If value of the x-coordinate is greater than the value of the y-coordinate, then the h-index is (x-1).
   c. If value of the x-coordinate is lesser than the value of the y-coordinate, then the h-index is x, which is never possible since Journal Number serial line is increasing and positively sloping.

ii. When the two lines, i.e., Journal Number serial line and Number of Citations line intersect each the value of the coordinates of the points of intersection both are not integers:
   a. If value of the x-coordinate equals to the value of the y-coordinate, then the h-index is equal to that integer, i.e., h-index = Floor of x, i.e.,
      $\lfloor x \rfloor$ = Floor of y, i.e., $\lfloor y \rfloor$
   b. If value of the x-coordinate is greater than the value of the y-coordinate, then the h-index is $\lfloor (x-1) \rfloor$.

iii. When the two lines, i.e., Journal Number serial line and Number of Citations line do not intersect each other the minimum value of the Euclidean/perpendicular distance of the corresponding points are calculated. Let the minimum value be m.
   a. The h-index is given by the x or y coordinate value of Journal Number serial line if the same equals to corresponding y – coordinate value of Citations line which is corresponding to m.
   b. The h-index is given by the (x-1), i.e. one less than the value of Journal Number serial corresponding to m if corresponding y – coordinate value of Citations line is less than the corresponding Journal Number serial.

------------------------------------------------------------